\documentclass[twocolumn]{revtex4-1}
\usepackage{graphics,graphicx,epsfig}

\begin{document}

\title{Competition among reputations in the 2D Sznajd model: Spontaneous emergence of democratic states}

\author{Nuno Crokidakis$^{1,2}$}
\email{nuno@if.uff.br}

\author{Fabricio L. Forgerini$^{2,3}$}
\email{fabricio$\_$forgerini@ufam.edu.br}

\affiliation{
$^{1}$ Instituto de F\'{\i}sica - Universidade Federal Fluminense \\
Av. Litor\^anea s/n, 24210-340 \hspace{2mm} Niter\'oi - Rio de Janeiro, Brazil \\
$^{2}$ Departamento de F\'{\i}sica, I3N - Universidade de Aveiro, 3810-193 \hspace{2mm} Aveiro, Portugal \\
$^{3}$ ISB, Universidade Federal do Amazonas, 69460-000 \hspace{2mm} Coari-AM, Brazil}

\date{\today}
\begin{abstract}
\noindent

We propose a modification in the Sznajd sociophysics model defined on the square lattice. For
this purpose, we consider reputation—a mechanism limiting the agents’ persuasive power. The reputation is introduced as a time-dependent score, which can be positive or negative. This mechanism avoids dictatorship (full consensus, all spins parallel) for a wide range of model parameters. We consider two different situations: case 1, in which the agents’ reputation increases for each persuaded neighbor, and case 2, in which the agents’ reputation increases for each persuasion and decreases when a neighbor keeps his opinion. Our results show that the introduction of reputation avoids full consensus even for initial densities of up spins greater than $1/2$. The relaxation times follow a log-normal-like distribution in both cases, but they are larger in case 2 due to the competition among reputations. In addition, we show that the usual phase transition occurs and depends on the initial concentration $d$ of individuals with the same opinion, but the critical points $d_{c}$ in the two cases are different.

\vspace{5mm}

\end{abstract}

\maketitle

%%%%%%%%%%%%%%%%%%%%%%%%%%%%%%%%%%%%%%%%%%%%%%%%%%%%%%%%%%%%%%%%%%%%%%%%%%%%%%

\section{Introduction}

\quad Ising-type models have been reviewed and used by physicists in many different areas, such as sociology, politics, marketing, and finance \cite{pmco_book, stauffer_review, sznajd_review, adriano}. In 2000, an agent-based model proposed by Sznajd-Weron \textit{et al.} \cite{sznajd_model} was successfully applied to the dynamics of a social system. In particular, the model reproduced certain properties observed in a real community. At the focus of the Sznajd model (SM) is the emergence of social collective (macroscopic) behavior due to the interactions among individuals, which constitute the microscopic level of a social system.

This model has been extensively studied since the introduction of the original one-dimensional model in 2000. Modifications were proposed in many works, like the consideration of different types of lattices such as square \cite{adriano}, triangular \cite{chang} and cubic \cite{bernardes}, the increase of the range of the interaction \cite{schulze1} and the number of variable's states \cite{bonnekoh,sznajds,stauffer_adv} and the possibility of diffusion of agents \cite{stauffer_adv,schulze2}.

The original SM consists of a chain of sites with periodic boundary conditions where each site (individual opinion) could have two possible states (opinions) represented in the model by Ising spins (``yes'' or ``no''). A pair of parallel spins on sites $i$ and $i+1$ forces its two neighbors, $i-1$ and $i+2$, to have the same orientation (opinion), while for an antiparallel pair $(i,i+1)$, the left neighbor ($i-1$) takes the opinion of the spin $i+1$ and the right neighbor ($i+2$) takes the opinion of the spin $i$. In this first formulation of the SM two types of steady states are always reached: complete consensus (ferromagnetic state) or stalemate (anti-ferromagnetic state), in which every site has an opinion that is different from the opinion of its neighbors. However, the transient displays a interesting behavior, as pointed by Stauffer et al. \cite{adriano}. Defining the model on a square lattice, the authors in \cite{adriano} considered not a pair of neighbors, but a $2\times 2$ plaquette with four neighbors. Considering that each plaquette with all spins parallel can convince all their eight neighbors (we will call this Stauffer's rule), a phase transition was found for an initial density of up spins $d=1/2$. 

It is more realistic to associate a probability of persuasion to each site. The SM is robust with respect to this choice: if one convinces the neighbors only with some probability $p$, and leaves them unchanged with probability $1-p$, still a consensus is reached after a long time \cite{stauffer_review}. Models that consider many different opinions (using Potts' spins, for example) or defined on small-world networks were studied in order to represent better approximations of real communities' behavior (see \cite{stauffer_review} and references therein). In another work, in order to avoid full consensus in the system and makes the model more realistic, Schneider introduced opportunists and persons in opposition, that are unconvinced by their neighbors \cite{schneider}. 

In the real world, however, the dynamics of social relationships is more complex. Even when such
more structured topologies as small-world networks are adopted to bring the SM closer to reality, a large number of details is often neglected. In order to advance toward realism, we recently considered a reputation mechanism \cite{sznajd_nosso}. We believe that the reputation
of agents who hold the same opinion is an important factor in opinion propagation across the community. In other words, it is realistic to believe that the individuals will change their opinions under the influence of highly respected persons. The reputation limits the agents’
power of persuasion, and we can expect the model in \cite{sznajd_nosso} to be more realistic than the standard one \cite{adriano}. In fact, we showed that simple microscopic rules are
sufficient to generate a democracy-like state, ferromagnetically ordered with only partial polarization \cite{sznajd_nosso}.

In this work, we revise and extend our previous results by allowing reputations to increase and decrease, depending on whether the agents are or are not persuaded. This generalization is based on the behavior of real social networks: Certain persons tend to be skeptical if the persuaders have low reputation, in which case their best strategy is to keep their opinions. In this
sense, including reputation makes the SM more realistic. To be thorough, we will consider two different protocols. In the first case, the agents' reputations increase for each persuaded neighbor, whereas in the second case, the agents' reputations rise in case of persuasion and decrease whenever the agents fail to convince their neighbors.

This paper is organized as follows: in Section \ref{model} we present the model and define their microscopic rules. The numerical results as well as the finite-size scaling analysis are discussed in Section \ref{results}. Finally, in Section \ref{conclusions} we summarize our conclusions.

%%%%%%%%%%%%%%%%%%%%%%%%%%%%%%%%%%%%%%%%%%%%%%%%%%%%%%%%%%%%%%%%%%%%%%%%%%%%%%%%

\section{Model}
\label{model}

\quad We have considered our model defined on a square lattice with $L \times L$ agents and periodic boundary conditions. Similar to Stauffer's rule (rule Ia of \cite{adriano}), we choose at random a $2 \times 2$ plaquette of four neighbors and if all central spins are parallel, the neighbors may change their opinions. The difference in our model is that the neighbor's spins will be flipped depending on the plaquette reputation. An integer number ($R$) labels each player and represents its reputation across the community, in analogy to the Naming game model considered by Brigatti \cite{edgardo}.
The reputation is introduced as a score for each player and is time dependent. The agents start with a random distribution of $R$ values, and during the time evolution, the reputation of each agent changes according to its capacity of persuasion. We will consider in this work that the initial values of the agents' reputation follow a gaussian distribution centered at $0$ with standard deviation $\sigma$.

We have considered two different situations in this work: in the first case, the reputations increase following the model's rules, whereas in the second case the reputations may increase and decrease. One time step in our model is defined by the following microscopic rules:\\

\vspace{0.3cm}
 \textit{Case 1}
\vspace{0.3cm}

\begin{enumerate}
\item We randomly choose a 2 $\times$ 2 plaquette of four neighbors;
\item If not all four center spins are parallel, leaves its eight neighbors unchanged.
\item On the other hand, if the four center spins are fully polarized, we calculate the average reputation $\bar{R}$ of the plaquette,
\begin{eqnarray}\nonumber
\bar{R} = \frac{1}{4}\sum_{i=1}^{4} R_{i}~,
\end{eqnarray}
where each term $R_{i}$ represents the reputation of one of plaquettes' agent.
\item We compare the reputations of each of the eight neighbors of the plaquette with the average reputation. If the reputation of a neighbor is less than the average, this neighbor follow the plaquette orientation. On the other hand, if the neighbor's reputation exceeds $\bar{R}$, no action is taken.
\item For each persuasion, the reputation of the plaquette agents is incremented by 1, so that the average plaquette reputation is increased by 1.
\end{enumerate}

\vspace{0.3cm}
 \textit{Case 2}
\vspace{0.3cm}

In this case, steps 1 - 4 are as described above. Step 5, by contrast, is changed to the following rule:

\begin{itemize}
\item For each persuasion, the reputation of the plaquette agents is incremented by 1. On the other hand, for each failure, the reputations within the plaquette are decremented by 1.

\end{itemize}

Thus, even in the case of fully polarized plaquettes, different numbers of agents may be convinced, namely 8, 7, 6, . . . , 1, or 0. As pointed by Stauffer in \cite{stauffer_review}, we
can imagine that each agent in the Sznajd model carries an opinion that can either be up (e.g., Republican) or down (e.g., Democrat), which represents one of two possible opinions on any question. The objective of the agents in the game is to convince their neighbors. One can expect that, if a certain group of agents convince many others, their persuasive power grows. On the
other hand, the persuasive powers may drop if the agents fail to convince other individuals. The inclusion of reputation in our model captures this feature of the real world.

%%%%%%%%%%%%%%%%%%%%%%%%%%%%%%%%%%%%%%%%%%%%%%%%%%%%%%%%%%%%%%%%%%%%%%%%%%%%%%

\section{Numerical Results}
\label{results}

%--------------------------------------------------------------------------------------------
\subsection{Case 1: Emergence of consensus}
\label{case1}

%%%%%%%%%%%%%%%%%%%%%%%%%%%%%%%%%%%%%%%%%%%%%%%%%%%%%%%%%%%%%%%%%%%%%%
\begin{figure}[t]
\begin{center}
\includegraphics[width=0.3\textwidth,angle=0]{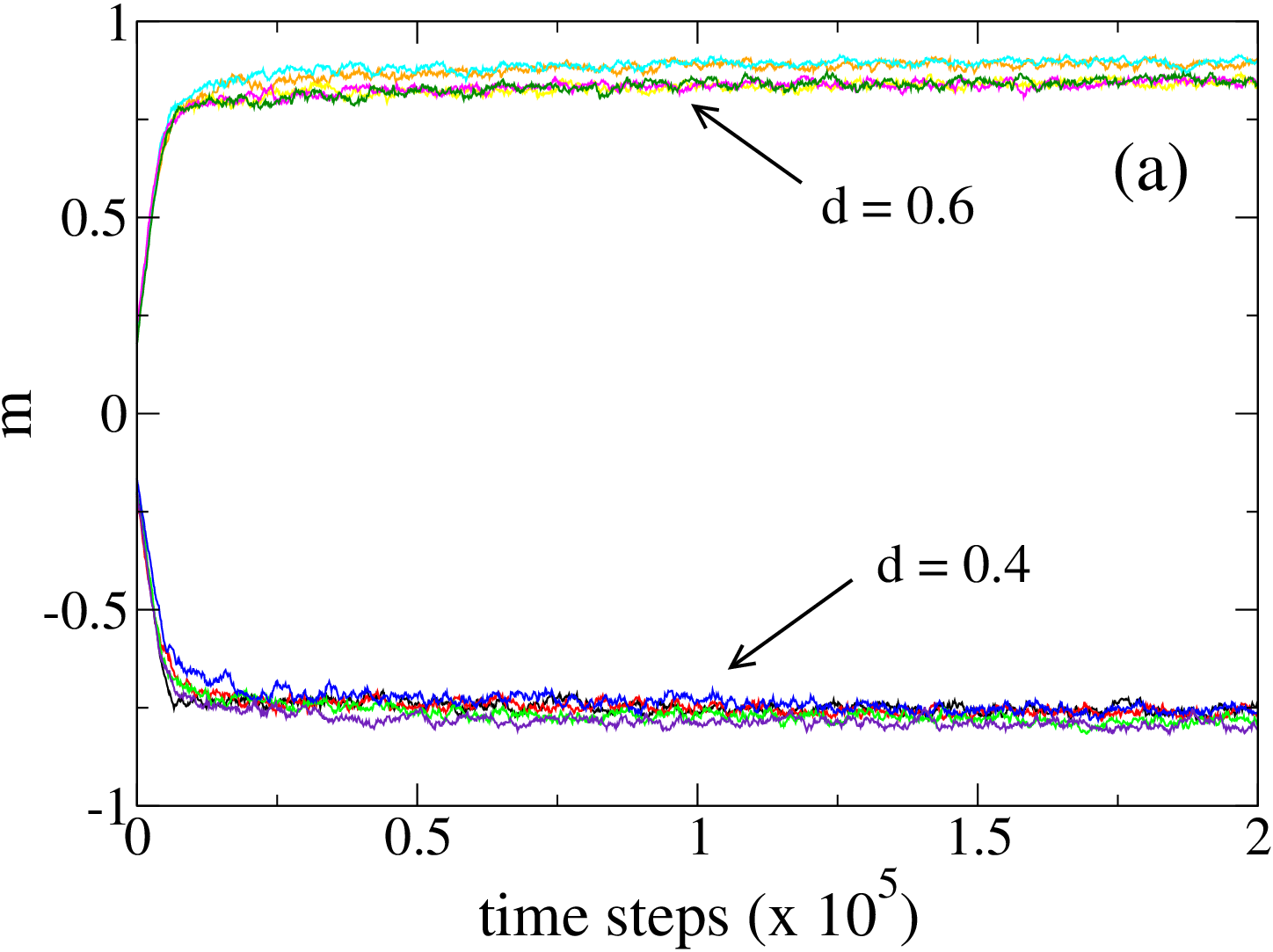}
\\
\vspace{0.7cm}
\includegraphics[width=0.3\textwidth,angle=0]{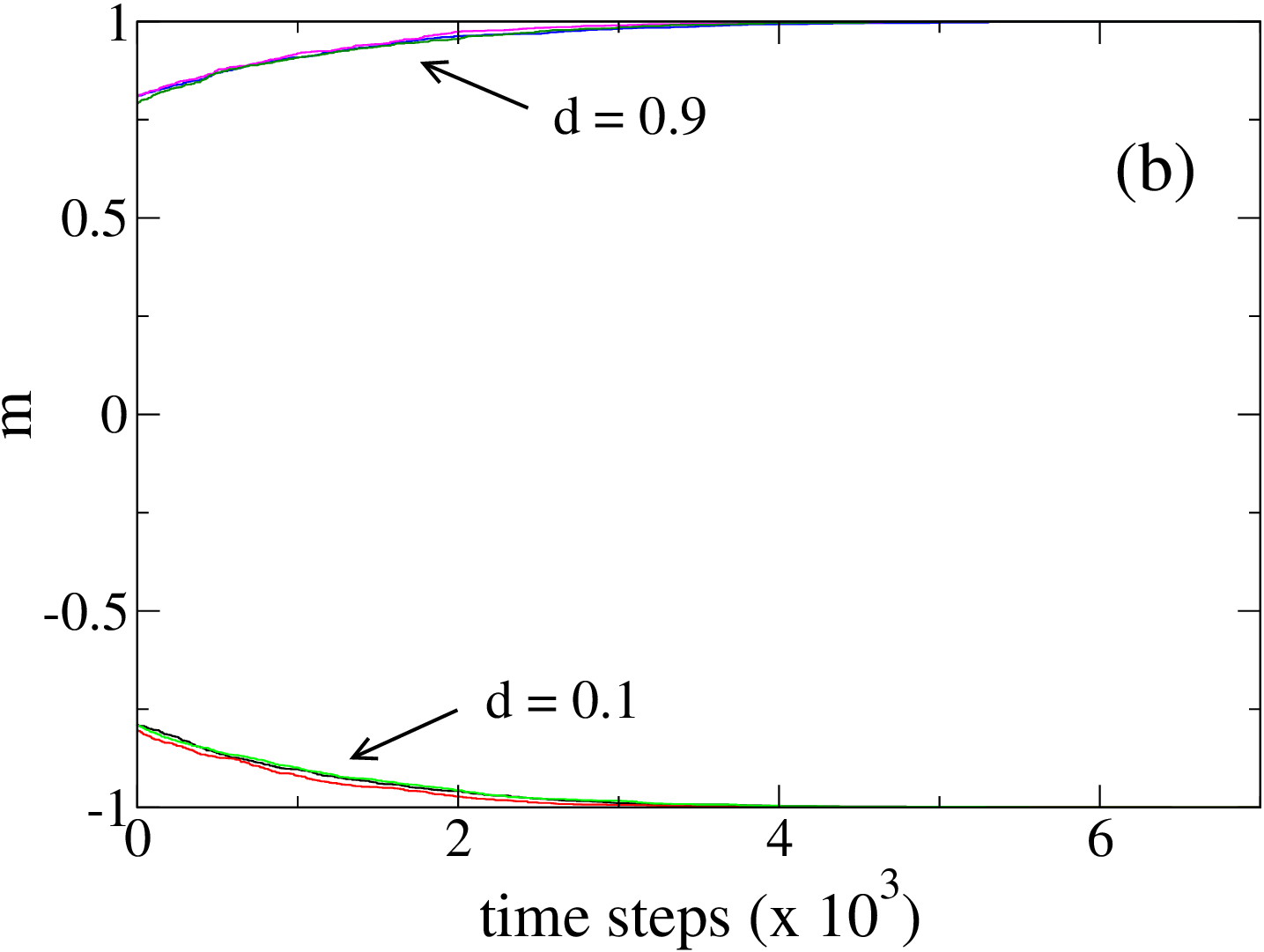}
\end{center}
\caption{Time evolution of the magnetization (case 1) for $L=53$, initial densities of up spins $d=0.4$ and $d=0.6$ and different samples (a). We can see that the steady states show situations where the total consensus is not obtained, in opposition of the standard Sznajd model defined on the square lattice \cite{adriano}. In figure (b) we show the results for $d=0.1$ and $d=0.9$. In these cases the system reaches consensus in all samples.}
\label{Fig1}
\end{figure}
%%%%%%%%%%%%%%%%%%%%%%%%%%%%%%%%%%%%%%%%%%%%%%%%%%%%%%%%%%%%%%%%%%%%%%

\quad In the simulations, we considered $\sigma=5$. Following the previous works on the SM, we can start studying the time evolution of the magnetization per site,
\begin{equation}
m=\frac{1}{N}\sum_{i=1}^{N}s_{i}~,
\end{equation}
\noindent
where $N=L^{2}$ is the total number of agents and $s_{i}=\pm 1$. In the standard SM defined on the square lattice \cite{adriano}, the application of the Stauffer's rule, where a $2\times 2$ plaquette with all spins parellel convince its eigth neighbors, with initial density of up spins $d=1/2$ leads the system to the fixed points with all up or all down spins with equal probability. For $d<1/2$ ($>1/2$) the system goes to a ferromagnetic state with all spins down (up) in all samples, which characterizes a phase transition at $d=1/2$ in the limit of large $L$. As pointed by the authors in \cite{adriano}, fixed points with all spins parallel describe the published opinion in a dictatorship, which is not a commom situation nowadays. However, ferromagnetism with not all spins parallel corresponds to a democracy, which is very commom in our world. We show in Fig. \ref{Fig1} the behavior of the magnetization as a function of the simulation time in our model, for case 1. In Fig. \ref{Fig1} (a), we show a value of $d>1/2$ ($<1/2$), and one can see that the total consensus with all spins up (down) will not be achieved in any sample. On the other hand, in Fig. \ref{Fig1} (b) we show situations where the consensus is obtained with all up (for $d=0.9$) and all down spins (for $d=0.1$). These results indicate that (i) a democracy-like situation is possible in the model without the consideration of a mixing of different rules \cite{adriano}, or some kind of special agents, like contrarians and opportunists \cite{schneider}, and (ii) if a phase transition also occurs in our case, the transition point will be located somewhere at $d>1/2$. 

%%%%%%%%%%%%%%%%%%%%%%%%%%%%%%%%%%%%%%%%%%%%%%%%%%%%%%%%%%%%%%%%%%%%%%
\begin{figure}[t]
\begin{center}
\includegraphics[width=0.3\textwidth,angle=0]{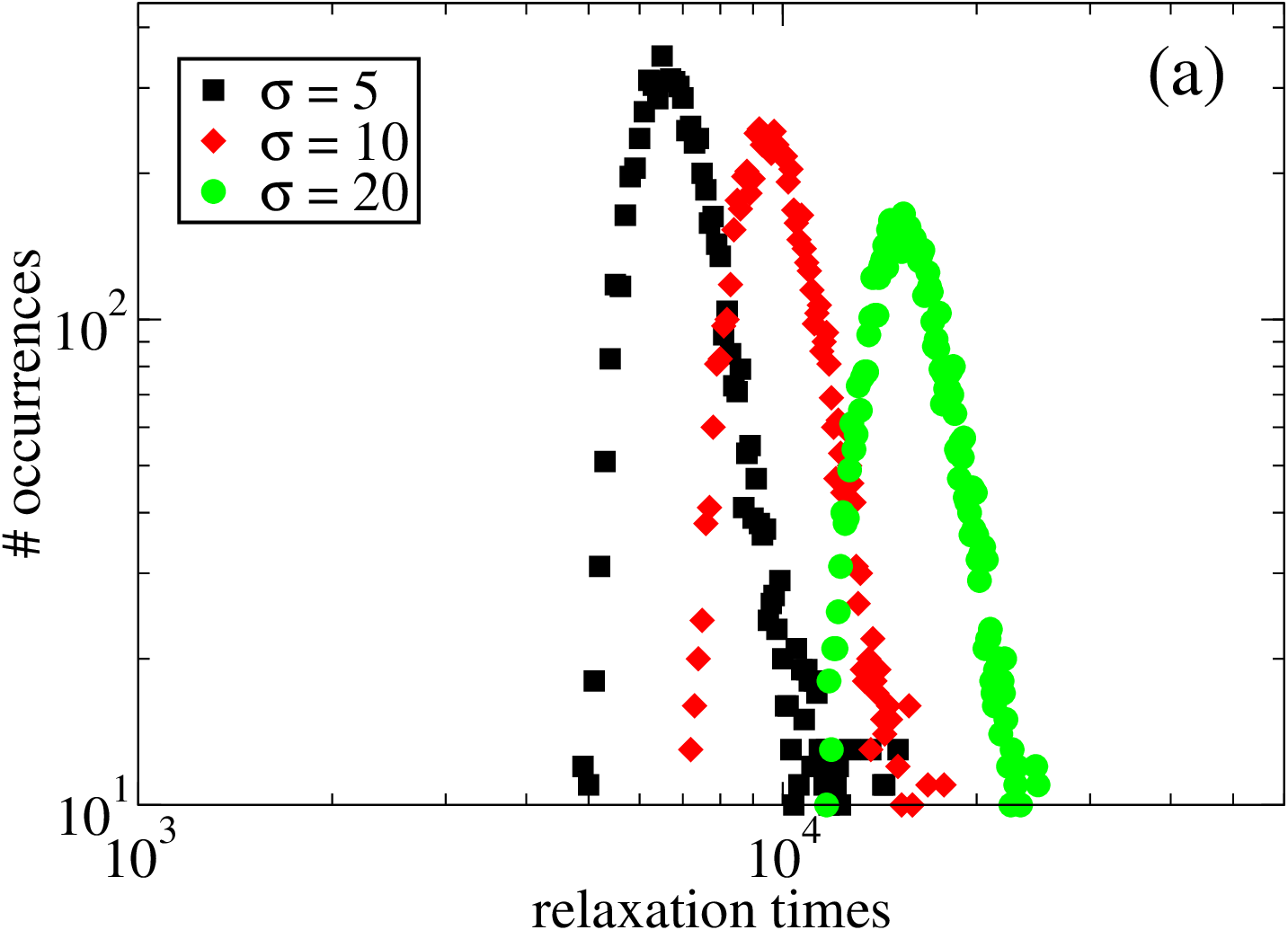}
\\
\vspace{0.7cm}
\includegraphics[width=0.3\textwidth,angle=0]{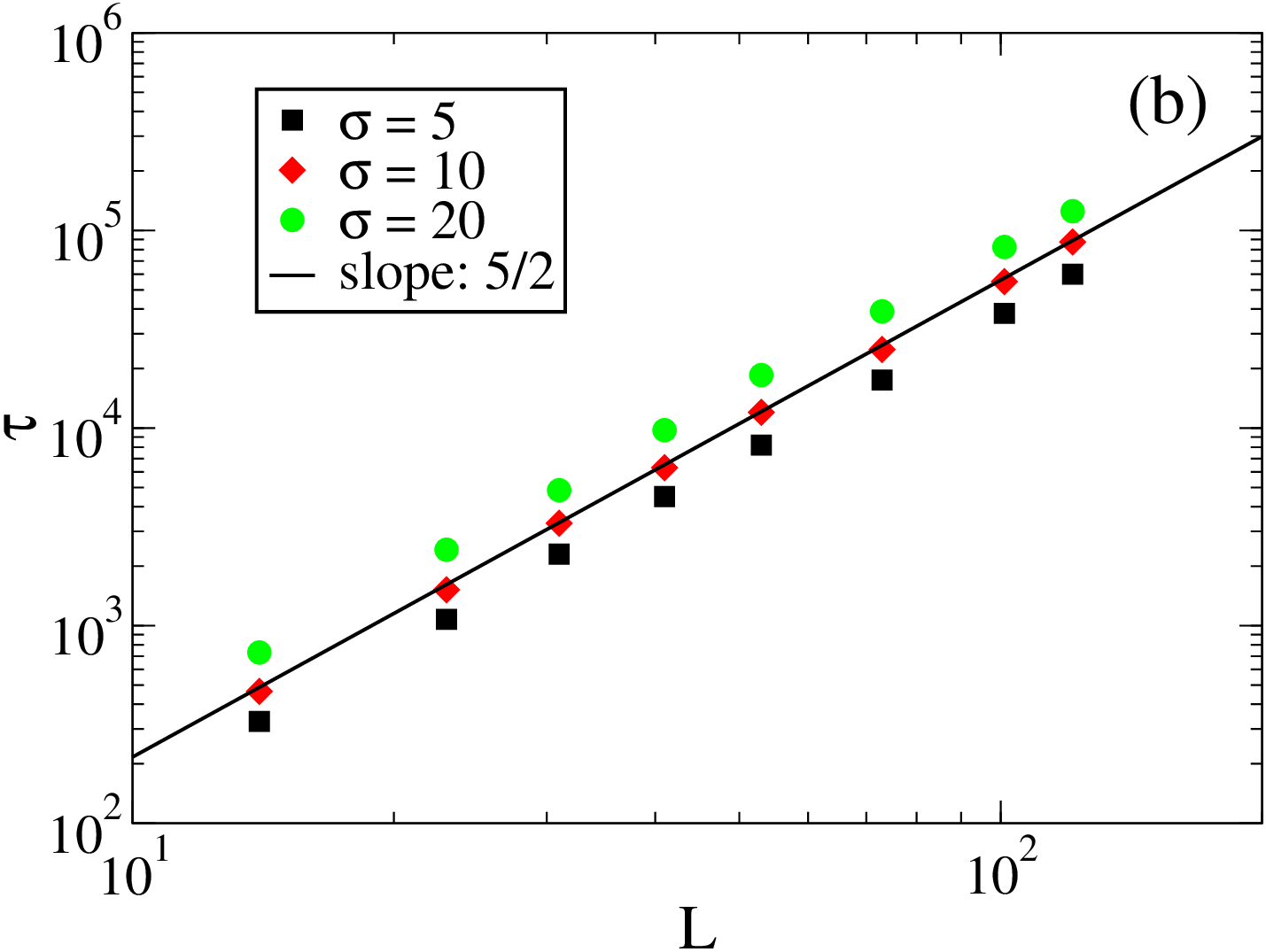}
\end{center}
\caption{Log-log plot of the histogram of relaxation times (case 1) for $L=53$ and $d=0.8$, obtained from $10^{4}$ samples, with agents' initial reputations following a gaussian distribution with different standard deviations $\sigma$ (a). The distribution is compatible with a log-normal one for all values of $\sigma$, which corresponds to the observed parabola in the log-log plot. It is also shown the average relaxation time $\tau$, over $10^{4}$ samples, versus latice size $L$ in the log-log scale (b). The straight line has slope 5/2. The result is robust with respect to the choice of different $\sigma$ values.}
\label{Fig2}
\end{figure}
%%%%%%%%%%%%%%%%%%%%%%%%%%%%%%%%%%%%%%%%%%%%%%%%%%%%%%%%%%%%%%%%%%%%%%

We have also studied the relaxation times of the model, i.e., the time needed to find all the agents at the end having the same opinion. The distribution of the number of sweeps through the lattice, averaged over $10^{4}$ samples, needed to reach the fixed point is shown in Fig. \ref{Fig2} (a). We can see that the relaxation time distribution is compatible with a log-normal one for all values of the standard deviation $\sigma$, which corresponds to a parabola in the log-log plot of Fig. \ref{Fig2} (a). The same behavior was observed in other studies of the SM \cite{adriano,schneider,adriano2}. In Fig. \ref{Fig2} (b) we show the average relaxation time $\tau$ [also over $10^{4}$ samples, considering the relaxation times of Fig. \ref{Fig2} (a)] versus latice size $L$ in the log-log scale. We can verify a power-law relation between these quantities in the form $\tau\sim L^{5/2}$, for large $L$ and all values of the standard deviation, which indicates that this result is robust with respect to the choice of different $\sigma$ values. Power-law relations between $\tau$ and $L$ were also found in a previous work on the SM \cite{adriano2}.

%%%%%%%%%%%%%%%%%%%%%%%%%%%%%%%%%%%%%%%%%%%%%%%%%%%%%%%%%%%%%%%%%%%%%%
\begin{figure}
\begin{center}
\includegraphics[width=0.3\textwidth,angle=0]{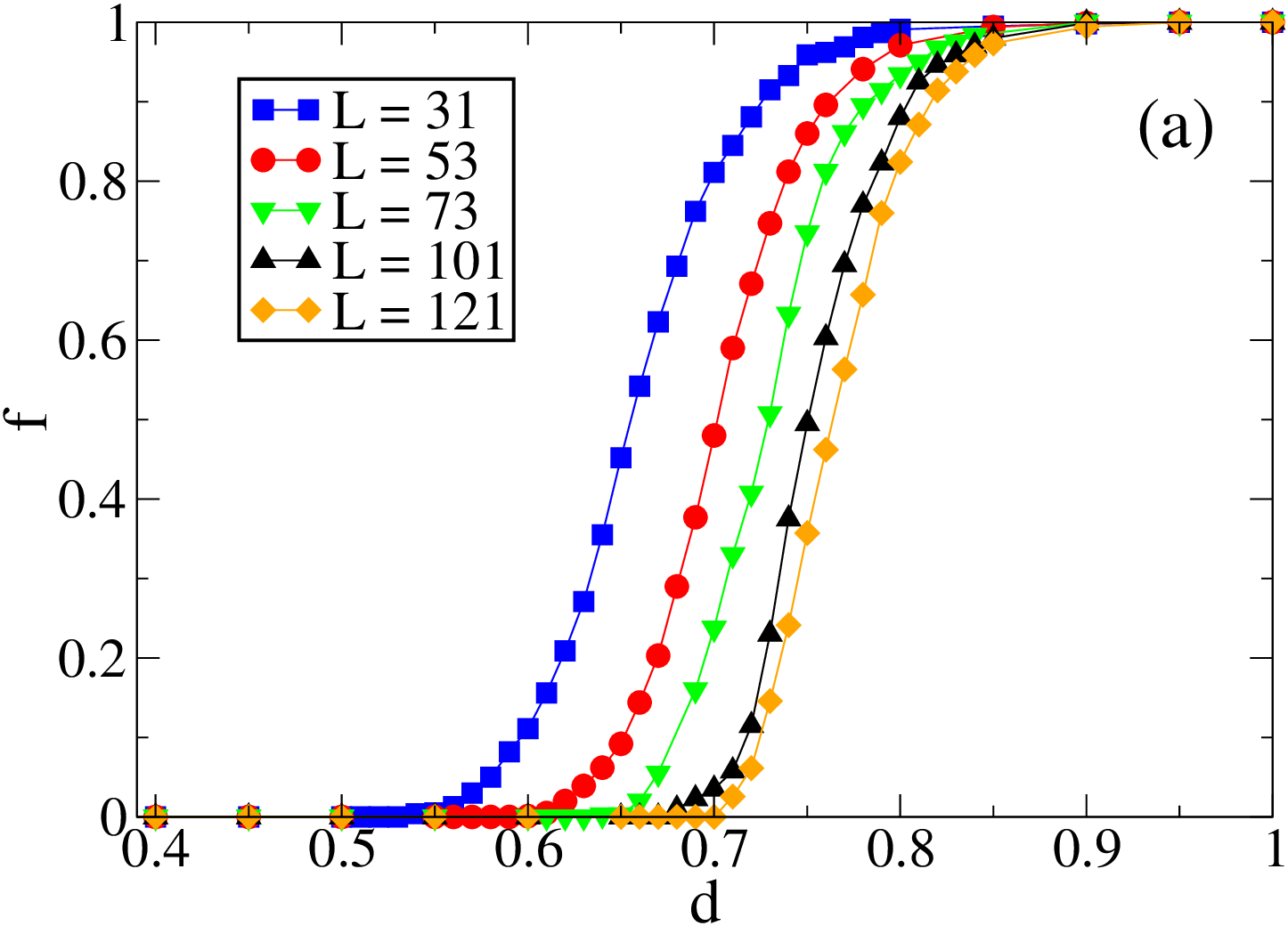}
\\
\vspace{0.7cm}
\includegraphics[width=0.3\textwidth,angle=0]{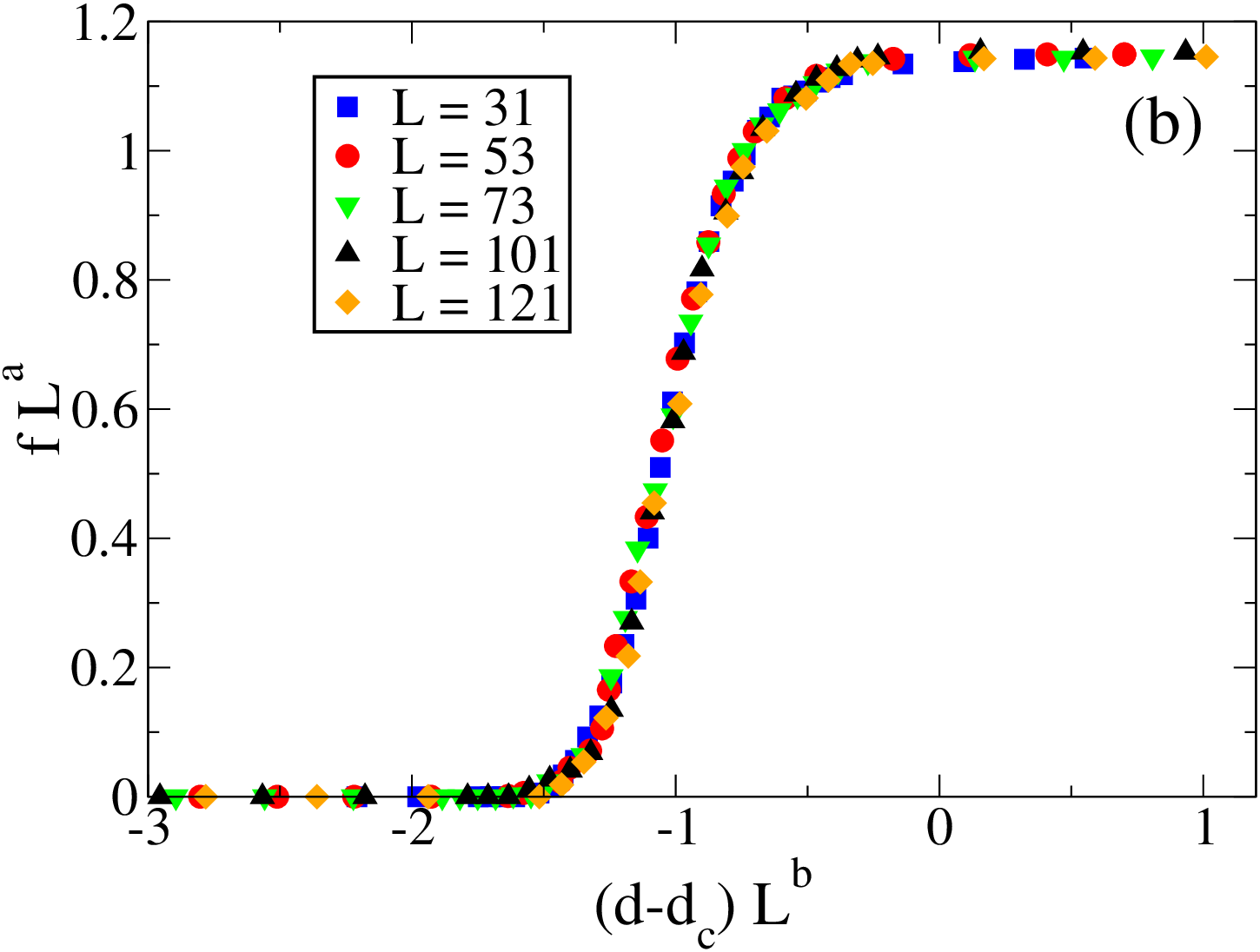}
\end{center}
\caption{Fraction $f$ of samples (case 1) which show all spins up when the initial density of up spins $d$ is varied in the range $0.4\leq d\leq 1.0$, for some lattice sizes $L$ (a). The total number of samples are $1000$ (for $L=31$ and $53$), $500$ (for $L=73$ and $101$) and $200$ (for $L=121$). It is also shown the corresponding scaling plot of $f$ (b). The best collapse of data was obtained for $a=0.03$, $b=0.47$ and $d_{c}=0.88$.}
\label{Fig3}
\end{figure}
%%%%%%%%%%%%%%%%%%%%%%%%%%%%%%%%%%%%%%%%%%%%%%%%%%%%%%%%%%%%%%%%%%%%%%

We can now analyze the phase transition of the model. For this purpose, we have simulated the system for different lattice sizes $L$ and we have measured the fraction of samples which show all spins up when the initial density of up spins $d$ is varied in the range $0.4\leq d\leq 1.0$. In other words, this quantity $f$ give us the probability that the population reaches consensus, for a given value of $d$. We have considered $1000$ samples for $L=31$ and $53$, $500$ samples for $L=73$ and $101$ and $200$ samples for $L=121$. The results are shown in Fig. \ref{Fig3} (a). One can see that the transition point is located somewhere in the region $d>1/2$, as above discussed. In order to locate the critical point, we performed a finite-size scaling (FSS) analysis, based on the standard FSS equations \cite{sznajd_nosso,adriano2},
\begin{eqnarray} \label{fss1}
f(d,L) & = & L^{-a}\;\tilde{f}((d-d_{c})\;L^{b}) ~, \\ \label{fss2}
d_{c}(L) & = & d_{c}+c\;L^{-b} ~,
\end{eqnarray}
\noindent
where $c$ is a constant and $\tilde{f}$ is a scaling function. The result is shown in Fig. \ref{Fig3} (b), and we have found that 
\begin{equation}
d_{c}=0.88 \pm 0.01~,
\end{equation}
%%%%%%%%%%%%%%%%%%%%%%%%%%%%%%%%%%%%%%%%%%%%%%%%%%%%%%%%%%%%%%%%%%%%%%
\begin{figure}[t]
\begin{center}
\includegraphics[width=0.3\textwidth,angle=0]{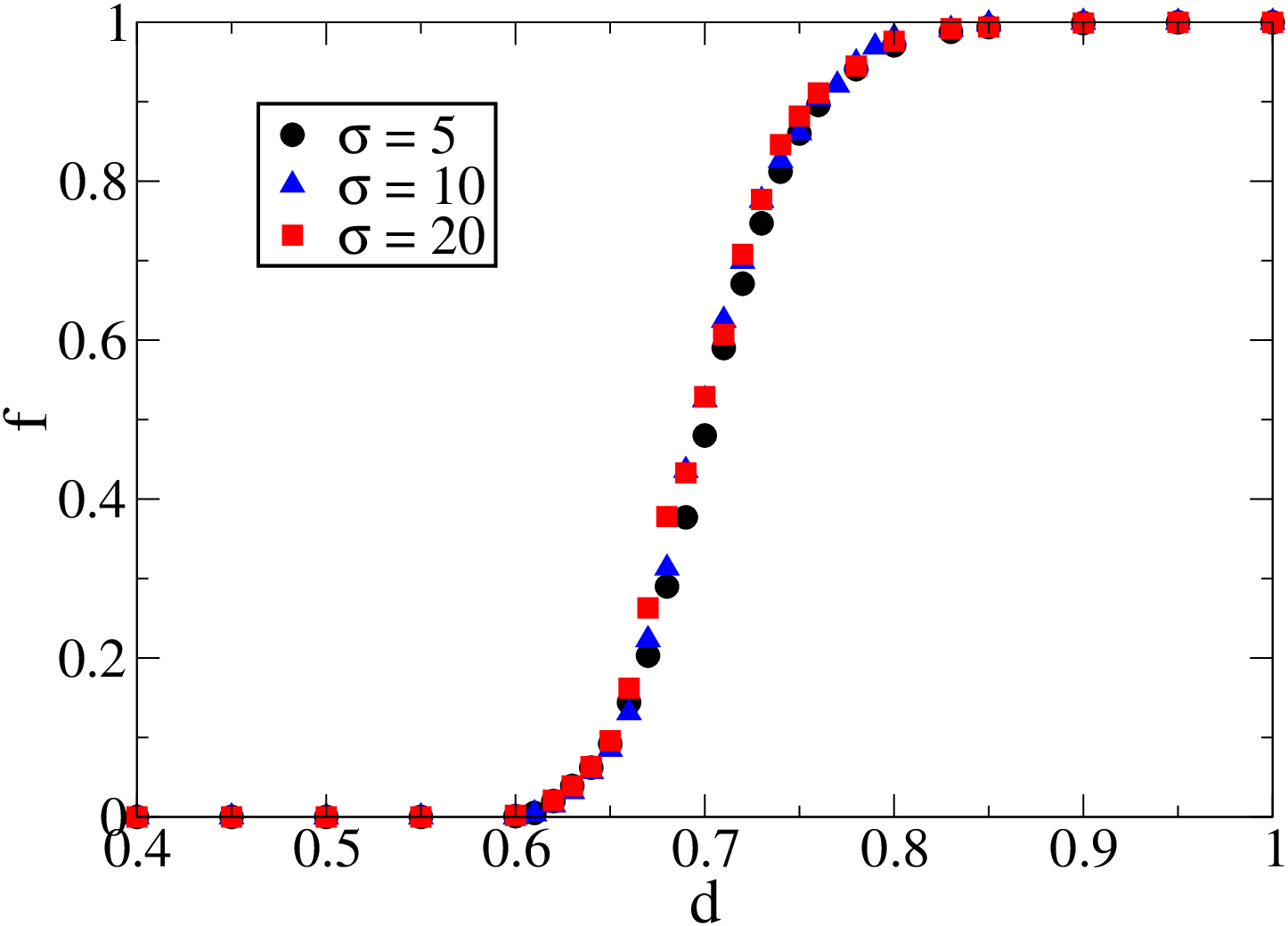}
\end{center}
\caption{Fraction $f$ of samples (case 1) which show all spins up when the initial density $d$ is varied in the range $0.4\leq d\leq 1.0$, for $L=53$, $1000$ samples and some different values of $\sigma$. This result show that the increase of $\sigma$ do not change the behavior of $f$.}
\label{Fig4}
\end{figure}
%%%%%%%%%%%%%%%%%%%%%%%%%%%%%%%%%%%%%%%%%%%%%%%%%%%%%%%%%%%%%%%%%%%%%%

\noindent
in the limit of large L. In addition, we have obtained $a=0.030\pm 0.005$ and $b=0.47\pm 0.02$. The critical point occurs at $d>1/2$, different of the SM without reputation defined on the square lattice. This fact may be easily understood: at each time step, the randomly choosen 2$\times$2 plaquette may convince 8, 7, 6, ..., 1 or 0 neighbors, even if the plaquettes' spins are parallel. In the standard model, if the plaquettes spins' orientations are the same, 8 neighbors are convinced immediately, thus it is necessary a smaller initial density of up spins to the system reaches the fixed point with all spins up. Thus, the usual phase transition of the SM also occurs in our model, in case 1, and this transition is robust with respect to the choice of different values of $\sigma$ (see Fig. \ref{Fig4}).

%--------------------------------------------------------------------------------------------
\subsection{Case 2: Competition among reputations}
\label{case2}

%%%%%%%%%%%%%%%%%%%%%%%%%%%%%%%%%%%%%%%%%%%%%%%%%%%%%%%%%%%%%%%%%%%%%%
\begin{figure}[t]
\begin{center}
\includegraphics[width=0.3\textwidth,angle=0]{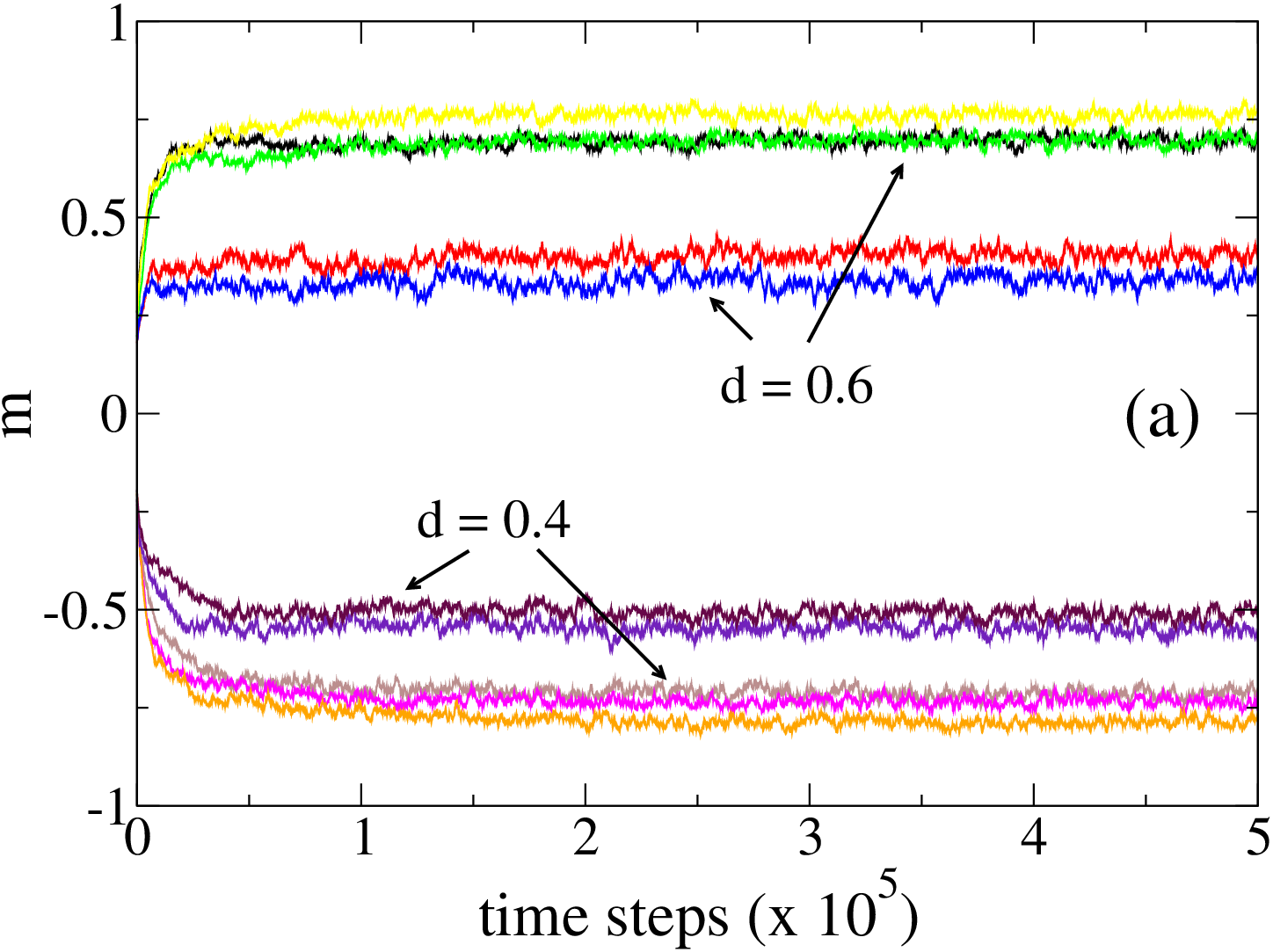}
\\
\vspace{0.7cm}
\includegraphics[width=0.3\textwidth,angle=0]{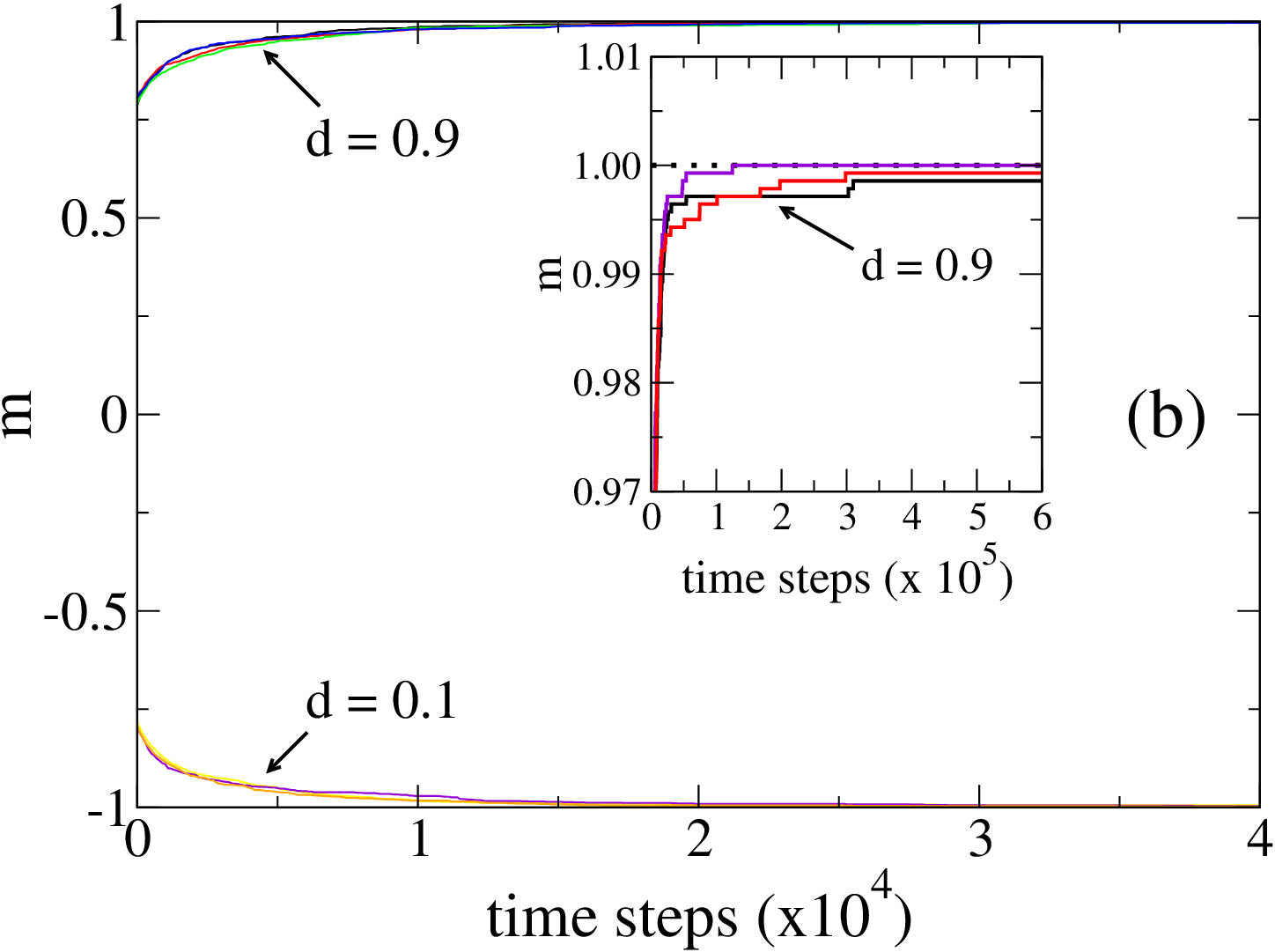}
\end{center}
\caption{Time evolution of the magnetization (case 2) for $L=53$, initial densities of up spins $d=0.4$ and $d=0.6$ and different samples (a). We can see differences between these steady states and those of case 1, but we also have democracy-like situations. In figure (b) we show the results for $d=0.1$ and $d=0.9$. Observe in the inset that even for large values of densities like $d=0.9$ the system reaches consensus only in some samples (analogously for $d=0.1$). The dotted line in the inset is $m=1$ (full consensus).}
\label{Fig5}
\end{figure}
%%%%%%%%%%%%%%%%%%%%%%%%%%%%%%%%%%%%%%%%%%%%%%%%%%%%%%%%%%%%%%%%%%%%%%

\quad As discussed in section \ref{model}, in this second case the agent's reputations may increase and decrease, which defines a competition of reputations in the game. The evolution of the magnetization per site is shown in Fig. \ref{Fig5}. In the case of intermediary densities $d$ the system reaches steady states with $m<1$, i.e., we have democracy-like situations. However, due to the competition of reputations, that increase and decrease depending on the average reputation of the plaquettes during the time evolution, the system reaches steady states with different magnetizations. Another consequence of the competition appears in the case of large and very small initial densities $d$: even for the cases $d=0.9$ and $d=0.1$ the system reaches consensus only in some realizations of the dynamics. This fact can be observed in the  inset of Fig. \ref{Fig5} (b): the dotted line is $m=1$, and we observe that just one of the three realizations reaches consensus. Thus, for case 2, the consensus is very hard to be obtained. Nonetheless, the emergence of democratic steady states is favoured in this second case, in comparison with case 1, which makes the model more realistic in this sense.

%%%%%%%%%%%%%%%%%%%%%%%%%%%%%%%%%%%%%%%%%%%%%%%%%%%%%%%%%%%%%%%%%%%%%%
\begin{figure}[t]
\begin{center}
%\vspace{0.6cm}
\includegraphics[width=0.3\textwidth,angle=0]{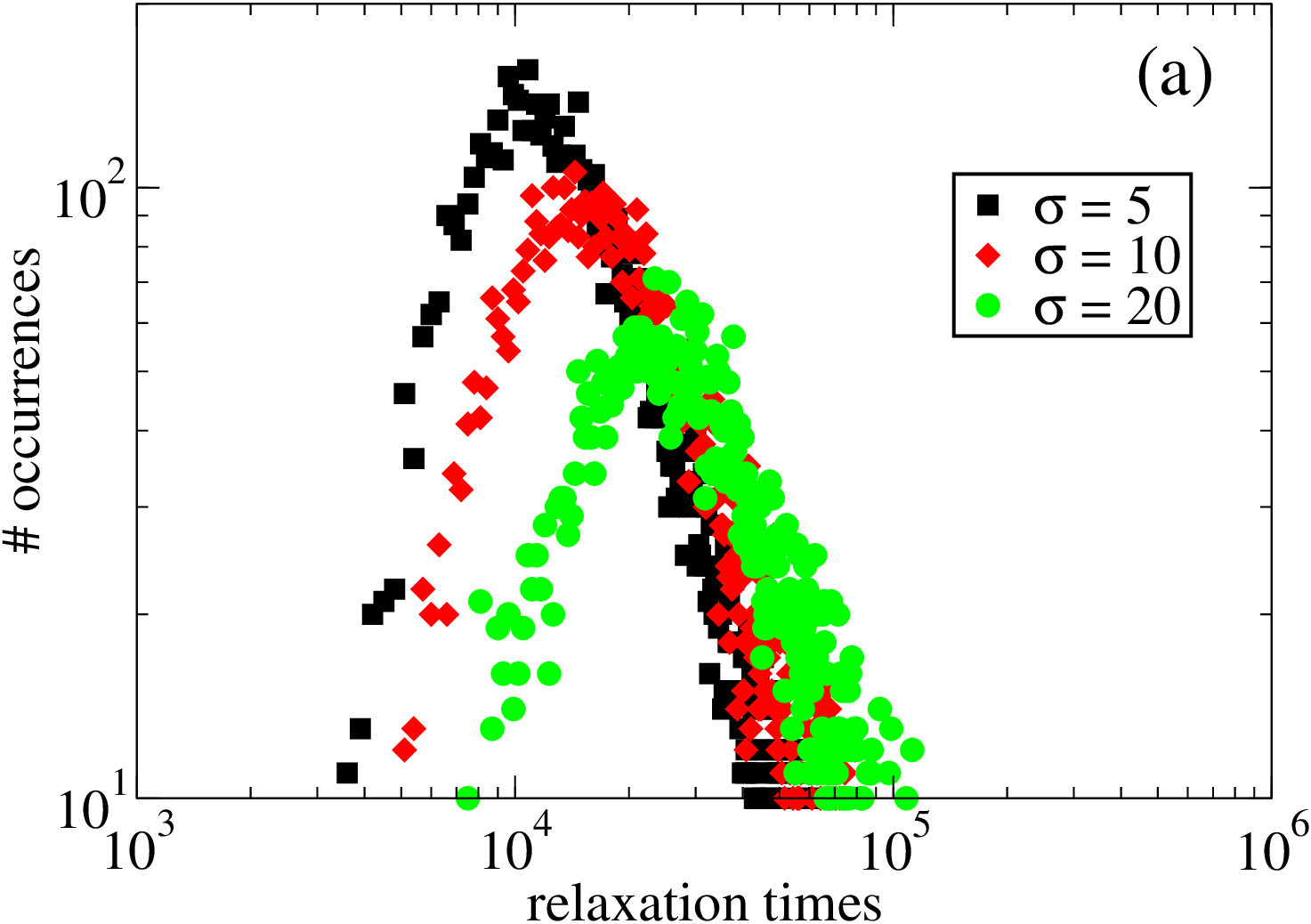}
\\
\vspace{0.7cm}
\includegraphics[width=0.3\textwidth,angle=0]{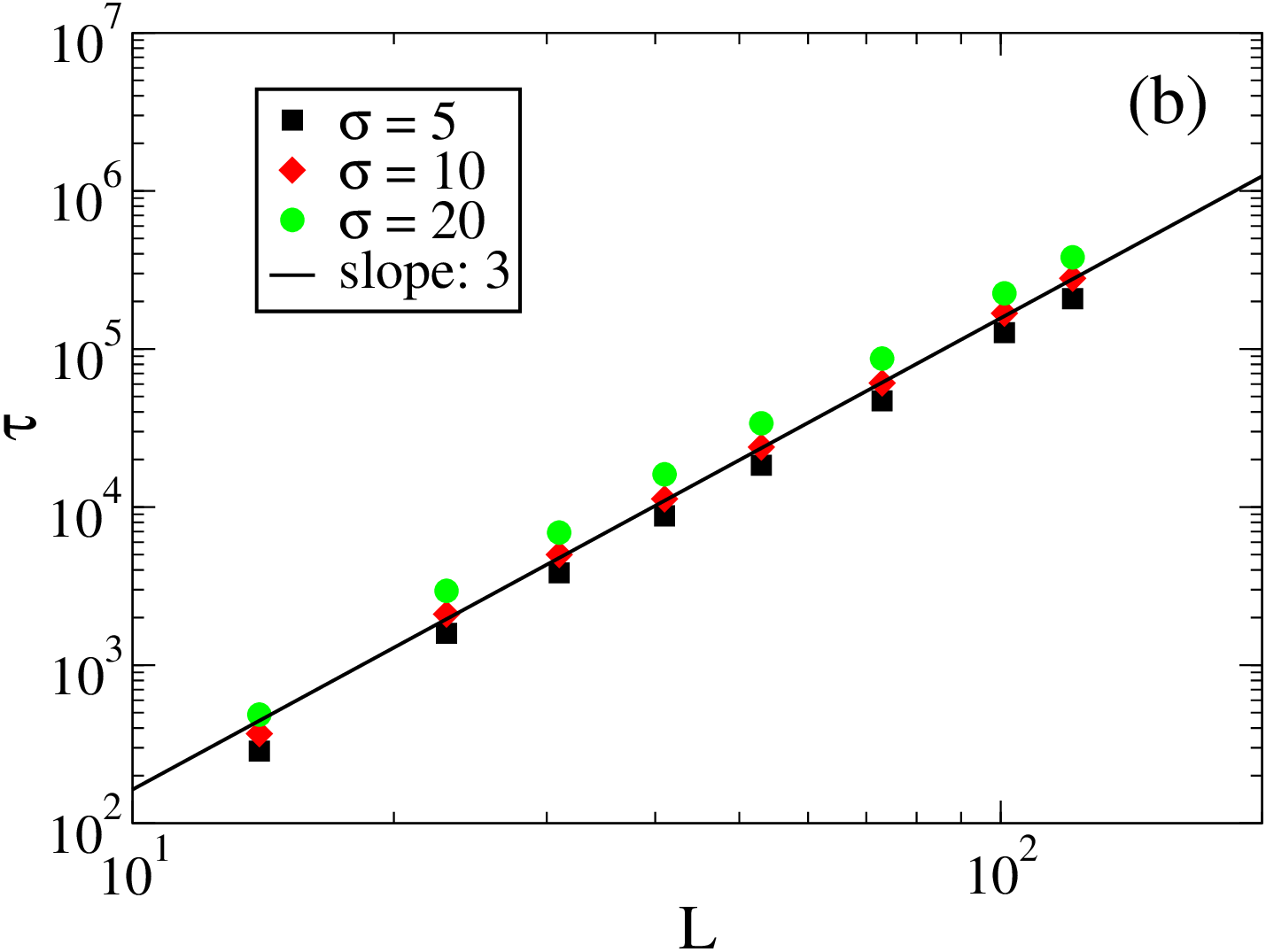}
\end{center}
\caption{Log-log plot of the histogram of relaxation times (case 2) for $L=53$ and $d=0.99$, obtained from $10^{4}$ samples, with agents' initial reputations following a gaussian distribution with different standard deviations $\sigma$ (a). The distribution is compatible with a log-normal one for all values of $\sigma$, which corresponds to the observed parabola in the log-log plot. It is also shown the average relaxation time $\tau$, over $10^{4}$ samples, versus latice size $L$ in the log-log scale (b). The power-law behavior for large $L$ is $\tau\sim L^{3}$, for all values of $\sigma$.}
\label{Fig6}
\end{figure}
%%%%%%%%%%%%%%%%%%%%%%%%%%%%%%%%%%%%%%%%%%%%%%%%%%%%%%%%%%%%%%%%%%%%%%

%%%%%%%%%%%%%%%%%%%%%%%%%%%%%%%%%%%%%%%%%%%%%%%%%%%%%%%%%%%%%%%%%%%%%%
\begin{figure}
\begin{center}
\includegraphics[width=0.3\textwidth,angle=0]{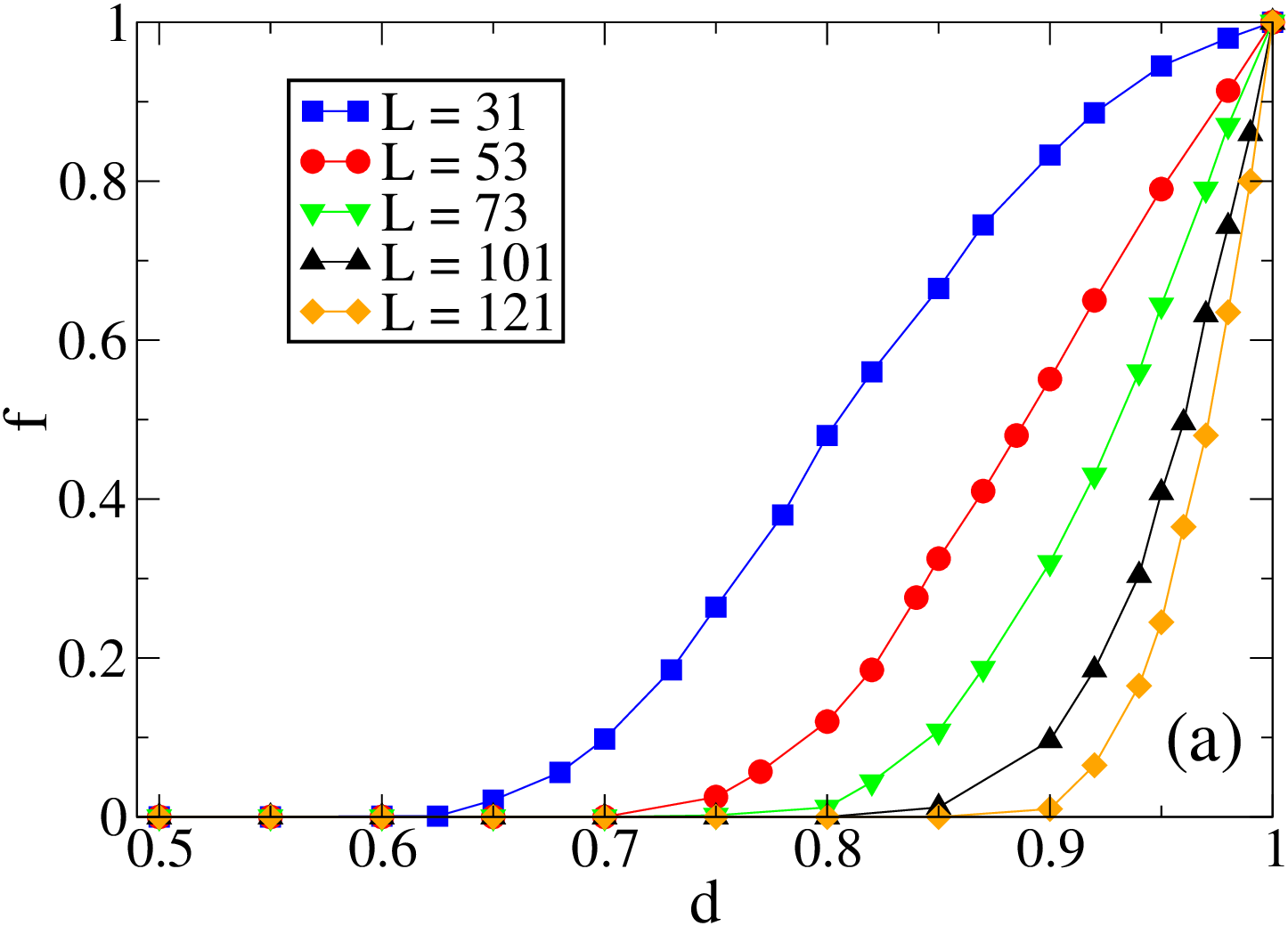}
\\
\vspace{0.7cm}
\includegraphics[width=0.3\textwidth,angle=0]{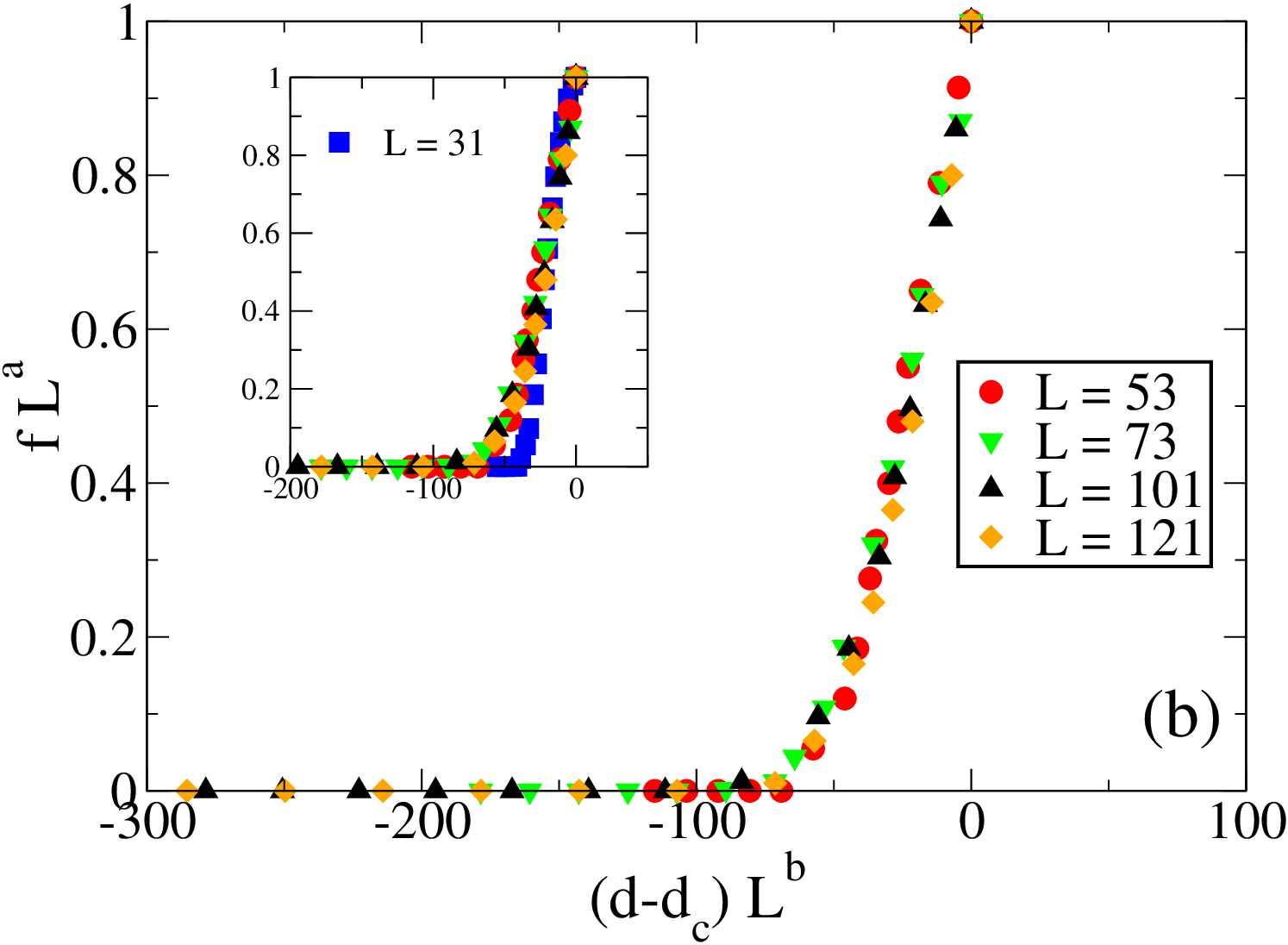}
\end{center}
\caption{Fraction $f$ of samples (case 2) which show all spins up when the initial density of up spins $d$ is varied in the range $0.5\leq d\leq 1.0$, for some lattice sizes $L$ (a). The total number of samples are $1000$ (for $L=31$ and $53$), $500$ (for $L=73$ and $101$) and $200$ (for $L=121$). It is also shown the corresponding scaling plot of $f$ (b). The best collapse of data was obtained for $a=0.0$, $b=1.37$ and $d_{c}=1.0$. To minimize the finite-size effects, we have excluded the smaller size $L=31$ of the collapse (see the inset).}
\label{Fig7}
\end{figure}
%%%%%%%%%%%%%%%%%%%%%%%%%%%%%%%%%%%%%%%%%%%%%%%%%%%%%%%%%%%%%%%%%%%%%%

We have also studied the relaxation times for case 2. The distribution of the number of sweeps through the lattice, averaged over $10^{4}$ samples, needed to reach the fixed point is shown in Fig. \ref{Fig6} (a) for some values of the standard deviation $\sigma$. We can see that, as in case 1, the relaxation time distribution is compatible with a log-normal one for all values of $\sigma$. However, due to competition of reputations, the relaxation times of case 2 are greater than the corresponding relaxation times of case 1. In Fig. \ref{Fig6} (b) we show the average relaxation time $\tau$ [also over $10^{4}$ samples, considering the relaxation times of Fig. \ref{Fig6} (a)] versus latice size $L$ in the log-log scale. In this case, we verify the power-law behavior $\tau\sim L^{3}$ for large $L$ and all values of the standard deviation. In other words, the competition of reputations increases the relaxation times of the system, as above discussed, and this effect becomes stronger when we increase the number of agents of the system (or the lattice size $L$), which implies in a power-law exponent for $\tau$ greater than the exponent for the case 1.

\begin{figure}[t]
\begin{center}
\vspace{0.6cm}
\includegraphics[width=0.3\textwidth,angle=0]{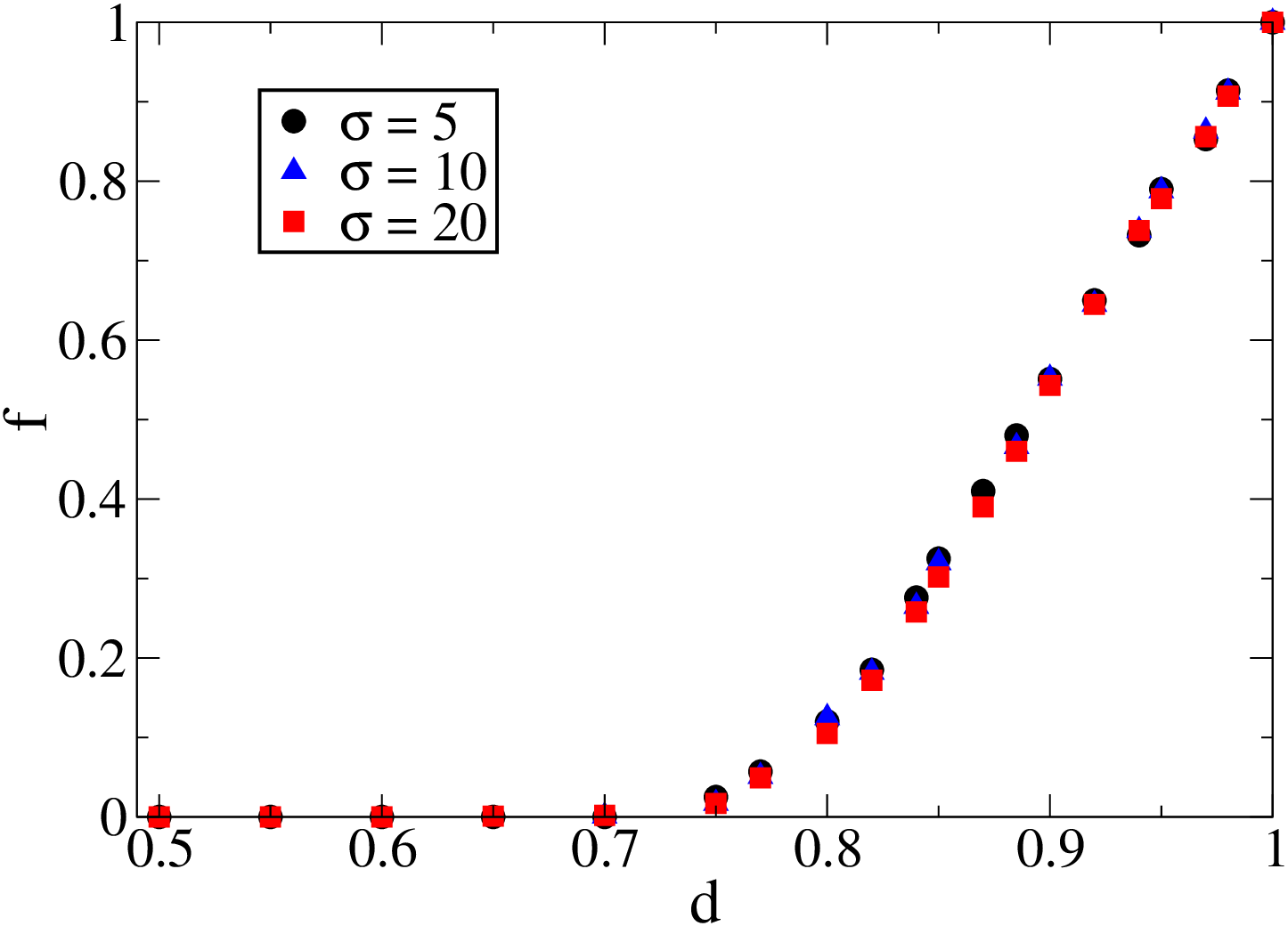}
\end{center}
\caption{Fraction $f$ of samples (case 2) which show all spins up when the initial density $d$ is varied in the range $0.5\leq d\leq 1.0$, for $L=53$, $1000$ samples and some different values of $\sigma$. This result show that the increase of $\sigma$ do not change the behavior of $f$.}
\label{Fig8}
\end{figure}

Following the approach of the last subsection (case 1), we have simulated the system for different lattice sizes $L$ and we have measured the fraction of samples which show all spins up when the initial density of up spins $d$ is varied in the range $0.5\leq d\leq 1.0$. We have considered the same number of samples of the last subsection, and the results are shown in Fig. \ref{Fig7} (a). One can see that the transition point is located somewhere in the region $d>0.88$, i.e., the critical density in case 2 is greater than in case 1, as expected due to the competition of reputations. We determined the critical point for this case using the above Eqs. (\ref{fss1}) and (\ref{fss2}). The best collapse of data is shown in Fig. \ref{Fig7} (b), obtained with $d_{c}=1.00\pm 0.01$, $a=0.00\pm 0.01$ and $b=1.37\pm 0.02$. In other words, the case 2 presents a different critical density and different critical exponents, in comparison with case 1. However, the usual phase transition of the SM also occurs in case 2, and this transition is robust with respect to the choice of different values of $\sigma$ (see Fig. \ref{Fig8}).

Observe that, in order to minimize the finite-size effects, we excluded the smaller size $L=31$ for the FSS process [see the inset of Fig. \ref{Fig7} (b)]. In fact, we can observe in Fig. \ref{Fig7} (a) that, for $L=31$, the curve of the quantity $f$ presents a inflection point, which not appears in the other sizes. Thus, this inflection point in the curve for $L=31$ is a pronounced finite-size effect of the model considered in case 2. 

%%%%%%%%%%%%%%%%%%%%%%%%%%%%%%%%%%%%%%%%%%%%%%%%%%%%%%%%%%%%%%%%%%%%%%
%%%%%%%%%%%%%%%%%%%%%%%%%%%%%%%%%%%%%%%%%%%%%%%%%%%%%%%%%%%%%%%%%%%%%%

%%%%%%%%%%%%%%%%%%%%%%%%%%%%%%%%%%%%%%%%%%%%%%%%%%%%%%%%%%%%%%%%%%%%%%%%%%%%%
\section{Conclusions}
\label{conclusions}

\quad In this work we have studied a modified version of the Sznajd sociophysics model. In particular we have considered reputation, a mechanism that limits the capacity of persuasion of the agents. The reputation is introduced as a score for each player and is time dependent, varying according to the model's rules. The agents start with a random distribution of reputation values, and during the time evolution, the reputation of each agent changes according to its capacity of persuasion. We have considered in this work that the initial values of the agents' reputation follow a gaussian distribution centered at $0$ with standard deviation $\sigma$. In addition, we have studied separately two different situations: (i) a case where the reputations increase due to each persuaded individual (case 1), and (ii) a case where the reputations increase for persuasion and decrease if a group of agents fail to convince one of its neighbors (case 2).

In the first case, we observed a log-normal-like distribution of the relaxation times, i.e., the time needed to find all the agents at the end having the same opinion. In addition, the average relaxation times grow with the linear dimension of the lattice in the form $\tau\sim L^{5/2}$. The system undergoes the usual phase transition, that was identified by measurements of the fraction $f$ of samples which show all spins up when the initial density of up spins $d$ is varied. In other words, this quantity $f$ give us the probability that the population reaches consensus, for a given value of $d$. We localized the transition point by means of a finite-size scaling analysis, and we found $d_{c}=0.88$. This critical density is greater than $1/2$, the value found by Stauffer \textit{et al.} \cite{adriano} in the standard formulation of the Sznajd model. This fact may be easily understood: at each time step, the randomly choosen 2$\times$2 plaquette may convince 8, 7, 6, ..., 1 or 0 neighbors, even if the plaquettes' spins are parallel. In the standard case, if the plaquettes spins' orientations are the same, 8 neighbors are convinced immediately, thus it is necessary a smaller initial density of up spins to the system reaches the fixed point with all spins up. The simulations indicate that the observed phase transition is robust with respect to the choice of different values of $\sigma$.

In the second case, the steady states with $m<1$ are favoured due to the competition of reputations, and even for large densities $d$ the system reaches consensus only in some samples. We also found that the relaxation times are log-normally distributed, but they are greater than the relaxation times of case 1. The average relaxation times are greater than the corresponding values found in the first case, and they also depend on the linear lattice size in a power-law form, $\tau\sim L^{3}$, but with a greater exponent in comparison with case 1. The usual phase transition also occurs in case 2, but the critical density was found to be $d_{c}=1.0$. In addition, the second situation presents strong finite-size effects. The observed differences between the two cases are due to the competition of reputations that occurs in case 2.

%%%%%%%%%%%%%%%%%%%%%%%%%%%%%%%%%%%%%%%%%%%%%%%%%%%%%%%%%%%%%%%%%%%%%%
\section*{Acknowledgments}

N. Crokidakis would like to thank the Brazilian funding agency CNPq for the financial support. Financial support from the Brazilian agency CAPES at Universidade de Aveiro at Portugal is also acknowledge. F. L. Forgerini would like to thank the ISB - Universidade Federal do Amazonas for the support.

%%%%%%%%%%%%%%%%%%%%%%%%%%%%%%%%%%%%%%%%%%%%%%%%%%%%%%%%%%%%%%%%%%%%%%%%%%%%%%%


\begin{thebibliography}{30}

\bibitem{pmco_book}
D. Stauffer, S. Moss de Oliveira, P. M. C. de Oliveira and J. S. S\'a Martins, \textit{Biology, Sociology, Geology by Computational Physicists} (Elsevier, Amsterdam, 2006).

\bibitem{stauffer_review}
D. Stauffer, J. Artif. Soc. Soc. Simul. \textbf{5}, 1 (2001). Available at
http://jasss.soc.surrey.ac.uk/5/1/4.html

\bibitem{sznajd_review}
K. Sznajd-Weron, Acta Phys. Pol. B \textbf{36}, 2537 (2005).

\bibitem{adriano}
D. Stauffer, A. O. Sousa and S. Moss de Oliveira, Int. J. Mod. Phys. C \textbf{11}, 1239 (2000).

\bibitem{sznajd_model}
K. Sznajd-Weron and J. Sznajd, Int. J. Mod. Phys. C \textbf{11}, 1157 (2000).

\bibitem{chang}
I. Chang, J. Mod. Phys. \textbf{12}, 1509 (2001).

\bibitem{bernardes}
A. T. Bernardes, D. Stauffer and J. Kertezs, Eur. Phys. J. B \textbf{25}, 123 (2002).

\bibitem{schulze1}
C. Schulze, Physica A \textbf{324}, 717 (2003).

\bibitem{bonnekoh}
J. Bonnekoh, Int. J. Mod. Phys. C \textbf{14}, 1231 (2003).

\bibitem{sznajds}
K. Sznajd-Weron and J. Sznajd, Physica A \textbf{351}, 593 (2005).

\bibitem{stauffer_adv}
D. Stauffer, Adv. Comp. Sys. \textbf{5}, 97 (2002).

\bibitem{schulze2}
C. Schulze, Int. J. Mod. Phys. C \textbf{14}, 95 (2003).

\bibitem{schneider}
J.J. Schneider, Int. J. Mod. Phys. C 15, 659 (2004).

\bibitem{sznajd_nosso}
N. Crokidakis and F. L. Forgerini, Phys. Lett. A \textbf{374}, 3380 (2010).

\bibitem{edgardo}
E. Brigatti, Phys. Rev. E \textbf{78}, 046108 (2008).

\bibitem{adriano2}
A. O. Sousa, T. Yu-Song and M. Ausloos, Eur. Phys. J. B \textbf{66}, 115 (2008).







\end{thebibliography}
\end{document}